\documentclass[12pt]{article}

\normalbaselines
\begin{document}
\begin{center}
\vspace*{1.0cm}

{\Large \bf Some Integrable Quantum Systems on the Lattice}

\vskip 1.5cm

{\large {\bf Miguel Lorente}}

\vskip 0.5 cm

Departamento de F\'{\i}sica (Facultad de Ciencias)\\
Universidad de Oviedo\\
33007 Oviedo, Spain

\end{center}

\vspace{1cm}

\begin{abstract}
The Weyl  relations, the harmonic  oscillator, the hydrogen atom, the Dirac equation
on the lattice are presented with the help of the difference equations and the
orthogonal polynomials of discrete variable. This area of research is attracting more
interest due to the lattice field theories and the hypothesis of a finite space.
\end{abstract}

\vspace{1cm}

\section{Weyl group on finite space}

We defined the position space of dimension N with orthonormal basis

$$\left| 0 \right\rangle =\left( {\matrix{1\cr 0\cr 0\cr
\vdots \cr 0\cr }} \right),\left| 1 \right\rangle =\left( {\matrix{0\cr 1\cr 0\cr
\vdots \cr 0\cr }} \right),\cdots ,\left| {N-1} \right\rangle =\left(
{\matrix{0\cr 0\cr 0\cr
\vdots \cr 1\cr }} \right),\left\langle {i} \mathrel{\left | {\vphantom {i j}}
\right. \kern-\nulldelimiterspace} {j} \right\rangle =\delta _{ij}$$

Similarly we construct an N-dimensional momentum space with ortonormal basis
$$\left| 0 \right\rangle =\left( {\matrix{1\cr
1\cr
1\cr
\vdots \cr
1\cr
}} \right),\left| 1 \right\rangle =\left( {\matrix{0\cr
\omega \cr
{\omega ^2}\cr
\vdots \cr
{\omega ^{N-1}}\cr
}} \right),\left| 2 \right\rangle =\left( {\matrix{1\cr
{\omega ^2}\cr
{\omega ^4}\cr
\vdots \cr
{\omega ^{2N-2}}\cr
}} \right),\left| {N-1} \right\rangle =\left( {\matrix{1\cr
{\omega ^{N-1}}\cr
{\omega ^{2N-2}}\cr
\vdots \cr
{{\omega ^{\left( {N-1} \right)}}^2}}} \right)$$
where $\omega ^N=1$, and $\left\langle {l} \mathrel{\left | {\vphantom {l k}} \right.
\kern-\nulldelimiterspace} {k}
\right\rangle =\delta _{lk}$

Two operators acting on these spaces are defined as
$$A=\left( {\matrix{0&1&0&0&\cdots \cr
0&0&1&0&\cdots \cr
0&0&0&1&\cdots \cr
\cdots &\cdots &\cdots &\cdots &\cdots \cr
1&0&0&0&\cdots \cr
}} \right),\;B=\left( {\matrix{1&{}&{}&{}&{}\cr
{}&\omega &{}&{}&{}\cr
{}&{}&{\omega ^2}&{}&{}\cr
{}&{}&{}&\ddots &{}\cr
{}&{}&{}&{}&{\omega ^{N-1}}\cr
}} \right),\;\;a,b\in Z$$
from which we construct $U_a\equiv A^a,\;V_b\equiv B^b,\;\;a,b\in Z$

On the position space, we have $U_a\left| j \right\rangle =\left| {j-a} \right\rangle ,\;V_b\left|
j \right\rangle =\omega ^{bj}\left| j \right\rangle $, (mod. $N$)
and in the momentum space
$U_a\left| k \right\rangle =\omega ^{ak}\left| k \right\rangle ,\;V_b\left| k \right\rangle
=\left| {k+b} \right\rangle $, (mod. $N$)

We can define the representation of vectors and operators as follows.

From the expansion in the momentum space

$$\left| F \right\rangle ={1 \over N}\sum\limits_{k=0}^{N-1} {\left| k \right\rangle \left\langle {k} \mathrel{\left | {\vphantom {k F}} \right. \kern-\nulldelimiterspace} {F} \right\rangle }=\sum\limits_{k=0}^{N-1} {b_k\left| k \right\rangle }$$
we obtain the representation in position space

\begin{equation} F\left( j \right)\equiv \left\langle {j} \mathrel{\left | {\vphantom {j F}}
\right.
\kern-\nulldelimiterspace} {F} \right\rangle =\sum {b_k}f_k\left( j \right)
\end{equation}
where$f_k\left( j \right)\equiv \left\langle {j} \mathrel{\left | {\vphantom {j k}} \right.
\kern-\nulldelimiterspace} {k} \right\rangle ={1 \over {\sqrt N}}\omega ^{jk}$

Similarly, from the expansion in momentum space,

$$\left| G \right\rangle ={1 \over N}\sum\limits_{k=0}^{N-1} {\left| j \right\rangle \left\langle {j} \mathrel{\left |
{\vphantom {j G}} \right. \kern-\nulldelimiterspace} {G} \right\rangle }=\sum\limits_{j=0}^{N-1} {a_j\left| j
\right\rangle }$$ we obtain the representation in momentum space
\begin{equation} G\left( k \right)\equiv \left\langle {k} \mathrel{\left | {\vphantom {j G}}
\right.
\kern-\nulldelimiterspace} {G} \right\rangle =\sum {a_j}f_j\left( k \right)
\end{equation}
For the operators $V_a$, $V_b$ we have in position space

$$\left( {U_aF} \right)\left( j \right)=\left\langle j \right|U_a\left| F \right\rangle =F\left( {j+a} \right)$$
$$\left( {V_bF} \right)\left( j \right)=\left\langle j \right|V_b\left| F \right\rangle =\omega ^{-bj}F\left( j \right)$$
and in momentum space

$$\left( {U_aG} \right)\left( k \right)=\left\langle k \right|U_a\left| G \right\rangle =\omega ^{-ak}G\left( k
\right)$$
$$\left( {V_bG} \right)\left( k \right)=\left\langle k \right|V_b\left| G \right\rangle =G\left( {k+b} \right)$$

From (1) and (2) we construct a Finite Fourier transform

$$\hat F\left( k \right)={1 \over {\sqrt N}}\sum\limits_{j=0}^{N-1} {F_j\omega ^{jk},\quad F_j=\left\langle {j} \mathrel{\left | {\vphantom {j F}} \right. \kern-\nulldelimiterspace} {F} \right\rangle
}$$
$$F\left( j \right)={1 \over {\sqrt N}}\sum\limits_{k=0}^{N-1} {\hat F\left( k \right)\omega ^{-jk},\quad \hat F\left( k \right)=\left\langle {k} \mathrel{\left | {\vphantom {k F}} \right. \kern-\nulldelimiterspace} {F} \right\rangle }$$

The Weyl approach to Quantum Mechanics[1] is based in the properties of the operators $A, B$ when the N-dimensional
space becomes infinite.

Postulate I. There exist two parameter abelian group in an N-dimensional space whose elements $A$ and $B$ satisfy
$$AB=\omega BA\quad , \quad \omega^N=1$$
$$A^s\left| j \right\rangle=\left| j+s \right\rangle\quad B^t\left| j \right\rangle=\omega^{jt}\left| j \right\rangle$$

Postulate II. In the continuous limit $N\rightarrow \infty$ we can identify
$$\omega \equiv e^{i{{2\pi } \over N}} \longrightarrow e^{i\xi \eta },\quad \xi <<1,\;\eta <<1$$
$$A^s\equiv \left( {e^{i\xi P}} \right)^s \longrightarrow e^{i\sigma P},\quad \xi s\to \sigma $$
$$B^t\equiv \left( {e^{i\eta Q}} \right)^t \longrightarrow e^{i\tau Q},\quad \eta t\to \tau $$
$$\omega ^{st}\equiv e^{is\xi t\eta } \longrightarrow e^{i\sigma \tau }$$
$$A^sB^t=\omega ^{st}B^tA^s \longrightarrow e^{i\sigma P}e^{i\tau Q}=e^{i\sigma \tau }e^{i\tau Q}e^{i\sigma P}$$

The justification of Postulate II lyes in the fact that in the continuous case the action of the translation and
multiplication operators is
$$U_\sigma f\left( x \right)\equiv e^{i\sigma P}f\left( x \right)=f\left( {x+\sigma } \right)$$
$$V_\tau f\left( x \right)\equiv e^{i\tau Q}f\left( x \right)=e^{i\tau q}f\left( x \right)$$
which are equivalent to the relations of Postulate I. From these equations the interpretation of the operators $P$,
$Q$ is derived as the generators of the translations and multiplications operators.

\section{The harmonic oscillator on the lattice}

In the discrete case we take the Kravchuk polynomials
${k}_{n}^{\left({p}\right)}\left({x,N}\right)$ with $x=0,1,2, \cdots \\ {N-1}$, [2] and for the
normalized functions the Wigner functions, that appear in the representation of the rotation
group, ${d}_{mm'}^{j}\left({\beta }\right)$
\[{\left({-1}\right)}^{m-m'}{d}_{mm'}^{j}\left({\beta }\right)={d}_{n}^{-1}\sqrt {\rho
\left({x}\right)}{k}_{n}^{(p)}\left({x,N}\right)\]
with \qquad $N=2j$ \qquad $m=j-n$ \qquad $m'=j-x$ \qquad $p ={\sin}^{2}{\displaystyle\frac{\beta
}{2}}\ ,\ q ={\cos}^{2}{\displaystyle\frac{\beta}{2}}$

After substitution in the fundamental formulas for the orthogonal polinomials we get [3] for the creation and
annihilation operators

\begin{eqnarray*}
\lefteqn{\sqrt {pq\left({n+1}\right)\left({N-n}\right)}{d}_{j-n-1,\ j-x}^{j}\left({\beta
}\right) =}\\
& &=p\left({N-x-n}\right){d}_{j-n,\ j-x}^{j}\left({\beta }\right)+\sqrt
{pqx\left({N-x+1}\right)}{d}_{j-n,\ j-x+1}^{j}\\
\lefteqn{\sqrt {pqn\left({N-n+1}\right)}{d}_{j-n+1,\ j-x}^{j}\left({\beta
}\right)=}\\
& & = p\left({N-x-n}\right){d}_{j-n,\ j-x}^{j}\left({\beta
}\right)+\sqrt {pq\left({x+1}\right)\left({N-x}\right)}{d}_{j-n,\
j-x-1}^{j}\left({\beta }\right)
\end{eqnarray*}

The last equations can be written down in terms of the new parameters

\[
\begin{array}{l}{\frac{1}{2}}\sin \beta \sqrt
{\left({j+m}\right)\left({j-m+1}\right)}{d}_{m-1,m'}^{j}\left({\beta }\right) =
{\sin}^{2}{\frac{\beta}{2}}\left({m+m'}\right){d}_{m,m'}^{j}\left({\beta }\right)+
\bigskip\\
\qquad+{\frac{1}{2}}\sin\
\beta \sqrt {\left({ j-m'}\right)\left({j+m'+1}\right)}{d}_{m,m'+1}^{j}\left({\beta }\right)\bigskip\\
{\frac{1}{2}}\sin\ \beta \sqrt {\left({ j-m}\right)\left({ j+m+1}\right)}{ d}_{ m+1,m'}^{
j}\left({\beta }\right) ={\sin}^{2}{\frac{\beta }{
2}}\left({m+m'}\right){d}_{m,m'}^{j}\left({\beta }\right)+ \bigskip\\
\qquad +{\frac{1}{2}}\sin\ \beta \sqrt
{\left({ j+m'}\right)\left({ j-m' +1}\right)}{ d}_{ m,m' -1}^{ j}\left({\beta }\right)
\end{array}\]

The creation and annihilation operators are connected with the
raising  and lowering operators for the spherical harmonics ${Y}_{jm}$. In fact, from the connection
between ${d}_{mm'}^{j}$ and $Y_{jm}$, we get
$$A{Y}_{jm}={\frac{1}{\sqrt {2j}}}\sqrt
{\left({j-m}\right)\left({j+m+1}\right)}{Y}_{j,m+1}={\frac{1}{\sqrt {2j}}}J_{+}{Y}_{jm}$$
$${A}^{\dagger}{Y}_{jm}={\frac{1}{\sqrt {2j}}}\sqrt
{\left({j+m}\right)\left({j+m+1}\right)}{Y}_{j,m-1}={\frac{1}{\sqrt {2j}}}J_{-}{Y}_{jm}$$

In order to make more transparent the connection between the creation and annihilation operators with
the raising and lowering operators of the spherical harmonics, we take the commutation and
anticommutation relations of the former operators.
\[\left[{A,{A}^{\dagger}}\right]{d}_{mm'}^{j}\left({\beta
}\right)=\left({1-{\frac{n}{j}}}\right){d}_{mm'}^{j}\left({\beta }\right)\]
wich in the limit $j\ \rightarrow \ \infty$ goes to
$$\left[{a,{a}^{\dagger}}\right]{\psi }_{n}\left({s}\right)={\psi }_{n}\left({s}\right)$$

Similarly 
\[\left({A{A}^{\dagger}+{A}^{\dagger}A}\right){d}_{mm'}^{j}\left({\beta
}\right)=\left\{{\left({2n+1}\right)-{\frac{{n}^{2}}{j}}}\right\}{d}_{mm'}^{j}\left({\beta }\right)\]
which in the limit $j\rightarrow \infty$ goes to 
$$\left({a{a}^{\dagger}+ {a}^{\dagger}a}\right){\psi }_{n}\left({s}\right)=\left({2n+1}\right){\psi
}_{n}\left({s}\right)$$

If we multiply both sides by $\hbar \omega /2$ we obtain the eigenvalue equation for the hamiltonian.

The interpretation of this model can be taken from the quantum harmonic oscillator.

The energy levels are equally distant by the amount $\hbar \omega$ and are labelled by $n=0,1,2,\cdots
\infty$. In the quantum harmonic oscillator of discrete variable we have also the discrete eigenvalues of
the hamiltonian connected with the index $m=j-n$ of the Wigner function ${d}_{mm'}^{j}\left({\beta }\right)$.

These values are equally separeted but finite $\left({m=-j,\cdots +j}\right)$. Similarly the eigenvalues of
the position operator $A+A^+$ are also discrete and connected to the index $m'=j-x$ of the Wigner functions
but finite $\left({m'=-j,\cdots ,+j}\right)$. 

The integer numbers $x=0,1,\cdots 2j$ are related to the quantity $x=\alpha s$ where $s$ is the continuous
variable and $\alpha =\sqrt {M\omega /\hbar }$. Since $x$ is a pure number and $s$ has the dimension of a
length, the spacing of the one-dimensional lattice is equal to $1/\alpha =\sqrt {\hbar /M\omega }$.
Therefore the Planck's constant $\hbar$ play role with respect to discrete space similar to the role with
respect to discrete energy values.

\section{The Hidrogen atom in the lattice}

We start from the difference equation for the Meixner polynomials, the limit of which goes to the
Laguerre polynomials in the continuous case. For the normalized Meixner polynomilas $M_n^{\left(
\gamma  \right)}\left( x \right)=d_n^{-1}\sqrt {\rho \left( x \right)}m_n^\gamma \left( x \right)$ we
get the following difference equation [3]
\begin{eqnarray*}
\lefteqn{\sqrt {\mu \left( {\gamma +x} \right)\left( {x+1} \right)}M_n\left( {x+1} \right)+\sqrt {\mu x\left(
{x+\gamma -1}\right)}M_n\left( {x-1} \right) -}\\ 
& & -\left[ {\mu \left( {x+n+\gamma } \right)-n+x} \right]M_n\left( x
\right)=0
\end{eqnarray*}

 For the Hidrogen atom in the continuous case one takes as the solution of the reduced radical equation
the wave function
$$\psi _n^{2l+1}\left( s \right)=\sqrt {\rho _1\left( s \right)}L_{n-l-1}^{2l+1}\left( s \right),\quad \rho
_1\left( s \right)=s\rho \left( s \right)$$
where $\rho \left( s \right)$ is the weight function. In the discrete case we take the wave function
$$U_n\left( x \right)=d_n^{-1}\sqrt {\rho _1\left( x \right)}M_n^\gamma \left( x \right),\quad \rho _1=\mu \left(
{x+\gamma } \right)\rho \left( x \right)$$

The difference equation now reads:
$$\sqrt {{{\mu \left( {x+1} \right)} \over {x+\gamma +1}}}U_n\left( {x+1} \right)+\sqrt {{{\mu x} \over {x+\gamma
}}}U_n\left( {x-1} \right)-{{\mu \left( {x+\gamma } \right)+x} \over {x+\gamma }}U_n\left( x \right)=\left( {\mu
-1} \right)n{{U_n\left( x \right)} \over {x+\gamma }}$$

This equation is of Sturn-Liouville type, from which an orthogonality relation can be derived:
$$\sum\limits_{x=0}^\infty  {{{U_m\left( x \right)U_n\left( x \right)} \over {x+\gamma }}}=0,\quad {\rm if}\;
n\ne m$$

We can construct also raising and lowering operators for the normalized Meixner functions. We get
$$L^+U_n\left( x \right)=\sqrt {\mu \left( {\gamma +n} \right)\left( {n-1} \right)}U_{n+1}\left( x \right)=\mu
\left( {x+n+\gamma } \right)U_n\left( x \right)-\sqrt {\mu x\left( {x+\gamma } \right)}U_n\left( {x-1} \right)$$
$$L^-U_n\left( x \right)=\sqrt {\mu n\left( {n+\gamma -1} \right)}U_{n-1}=\mu \left( {x+\gamma +n}
\right)U_n\left( x \right)-\mu \left( {x+\gamma } \right)\sqrt {{{x+1} \over {x+\gamma +1}}}U_n\left( {x+1}
\right)$$

The action of these operators is to create or annihilate a new state the eigenvalue of which (with respect to the
energy operator) is increased or decreased by unity.

\section{Dirac and Klein-Gordon equation on the lattice}

From the Dirac equation in momentum space $\left( {\gamma _\mu p_\mu -m_0c} \right)\psi \left( p \right)=0$
 we can construct the wave equation in position space with the help of
the Fourier transform. We define the following difference operators
\[\Delta  f\left({j}\right)=f\left({j+1}\right)-f\left({j}\right),\ \ \tilde{\Delta
}f\left({j}\right)={\frac{1}{2}}\left\{{f\left({j+1}\right)+f\left({j}\right)}\right\},\]
\[\nabla  f\left({j}\right)=f\left({j}\right)-f\left({j-1}\right),\ \ \tilde{\nabla
}f\left({j}\right)={\frac{1}{2}}\left\{{f\left({j}\right)+f\left({j-1}\right)}\right\},\]
and the partial difference operators with respect to a function of several discrete variables
\[{\Delta }_{\nu }f\left({{j}_{\mu }}\right)=f\left({{j}_{\mu }+{\delta }_{\mu \nu }}\right)-f\left({{j}_{\mu }}\right),\]
\[
{\tilde{\Delta }}_{\nu }f\left({{j}_{\mu }}\right)={\frac{1}{2}}\left\{{f\left({{j}_{\mu }+{\delta }_{\mu \nu
}}\right)+f\left({{j}_{\mu }}\right)}\right\},\]
and similarly ${\nabla }_{\nu }f\left({{j}_{\mu }}\right)$ and ${\tilde{\nabla }}_{\nu }f\left({{j}_{\mu }}\right)$.

From these operators we construct
\[{\delta }_{\mu }^{+}\equiv {\frac{ 1}{\varepsilon }}{\Delta }_{\mu }\prod\limits_{\nu \ne \mu }^{} {\tilde{\Delta
}}_{\nu }\
 \ ,\ \ {\delta }_{\mu }^{-}\equiv {\frac{ 1}{\varepsilon }}{\nabla }_{\mu }\prod\limits_{\nu \ne \mu }^{} {\tilde{\nabla
}}_{\nu }\ ,\]
\[{\eta }^{+}\equiv \prod\limits_{\mu  =0}^{ 3} {\tilde{\Delta }}_{\mu }\  \ ,\ \ {\eta }^{-}\equiv \prod\limits_{\mu 
=0}^{ 3} {\tilde{\nabla }}_{\mu }\ .\]

From the Fourier transform we can derive the wave equation in lattice space.

The kernel of the transform satisfies:
\[{\frac{1}{\varepsilon }}{\Delta }_{\mu }\exp\ \left(2\pi  i\left({k.j}\right)\varepsilon\right) 
=i{\frac{2}{\varepsilon }}\tan\ \left(\pi { k}_{\mu }\varepsilon\right) {\tilde{\Delta }}_{\mu } \exp\ \left(2\pi 
i\left({k.j}\right)\varepsilon\right) \]

We could apply the Fourier transform to the Dirac equation in momentum space and we would obtain the discrete wave
equation.
 Instead we postulate a difference equation that in the limit goes to the continuous differential equation, namely,
\[\left({i{\gamma }^{\mu }{\delta }_{\mu }^{+}-{m}_{0}c{\eta }^{+}}\right)\psi \left({{ j}_{\mu }}\right)=0.\]

The kernel of the Fourier transform or ``plane wave'' is a particular solution of this equation if it satisfies
\[
\left({{\gamma }^{\mu }{\frac{2}{\varepsilon }}\tan\pi { k}_{\mu }\varepsilon  -{m}_{0}c}\right){\eta }^{+}\ \exp\ 2\pi
 i\left({k\cdot  j}\right)\varepsilon  =0.\]

Applying the operator ${\gamma }^{\mu }{\frac{2}{\varepsilon }}\tan\pi { k}_{\mu }\varepsilon  +{m}_{0}c$ from the left
to the last equation we obtain
\[{\frac{4}{{\varepsilon }^{2}}}\left({\tan\ \pi { k}^{\mu }\varepsilon }\right)\left({\tan\ \pi { k}_{\mu }\varepsilon
}\right)-{m}_{0}^{2}{c}^{2}=0,\]
which is the integrability condition for the solution of the wave equation.

Applying to the wave equation the operator $i{\gamma }^{\mu }{\delta }_{\mu }^{-}+{m}_{0}c{\eta }^{-}$ from the left we obtain
the discrete version of the Klein-Gordon equation in the lattice space
\[\left({{\delta }_{\mu }^{+}{\delta }^{\mu \rm -}-{m}_{0}^{2}{c}^{2}{\eta }^{+}{\eta }^{-}}\right)\psi \left({{j}_{\mu
}}\right)\rm =0\]
a particular solution of which is again the ``plane wave'' provided the integrability conditions is satisfied.
\section*{Acknowledgment(s)}
The autor expresses his gratitude to Prof. Doebner, Dobrev, Hennig and Luecke for the invitations to the Symposium.

This work has been partially supported by D.G.I.C.Y.T. (Spain) \# Pb 96 0538.

\end{document}